\documentclass[aps,nature,british,superscriptaddress,floatfix,letterpaper,nonbalancelastpage,reprint]{revtex4-1}
\usepackage{graphicx} 
\usepackage{babel}
\usepackage{hyperref} 
\usepackage{amsmath} 
\usepackage{amssymb}
\newcommand{\ket}[1]{\left\lvert #1 \right\rangle}

\newcommand\startsupplement{%
    \makeatletter 
       \setcounter{table}{0}
       \renewcommand{\thetable}{S\arabic\c@table}
       \setcounter{figure}{0}
       \renewcommand{\thefigure}{S\@arabic\c@figure}
    \makeatother}

\usepackage{newfloat}
\DeclareFloatingEnvironment[name={Extended Data Figure}]{suppfigure}

\begin{document}
\title{Detecting arbitrary quantum errors via stabilizer measurements on a sublattice of the surface code}

\author{A. D. C\'orcoles*}
\affiliation{IBM T.J. Watson Research Center, Yorktown Heights, NY 10598, USA}
\author{Easwar Magesan*}
\affiliation{IBM T.J. Watson Research Center, Yorktown Heights, NY 10598, USA}
\author{Srikanth J. Srinivasan*}
\affiliation{IBM T.J. Watson Research Center, Yorktown Heights, NY 10598, USA}
\author{Andrew W. Cross}
\affiliation{IBM T.J. Watson Research Center, Yorktown Heights, NY 10598, USA}
\author{M. Steffen}
\affiliation{IBM T.J. Watson Research Center, Yorktown Heights, NY 10598, USA}
\author{Jay M. Gambetta}
\affiliation{IBM T.J. Watson Research Center, Yorktown Heights, NY 10598, USA}
\author{Jerry M. Chow}
\affiliation{IBM T.J. Watson Research Center, Yorktown Heights, NY 10598, USA}

\date{\today}
\maketitle

\noindent\textbf{\boldmath To build a fault-tolerant quantum computer, it is necessary to implement a quantum error correcting code. Such codes rely on the ability to extract information about the quantum error syndrome while not destroying the quantum information encoded in the system. Stabilizer codes~\cite{gottesman} are attractive solutions to this problem, as they are analogous to classical linear codes\cite{ECC}, have simple and easily computed encoding networks, and allow efficient syndrome extraction. In these codes, syndrome extraction is performed via multi-qubit stabilizer measurements, which are bit and phase parity checks up to local operations. Previously, stabilizer codes have been realized in nuclei~\cite{Cory:1998oq,PhysRevLett.109.100503,PhysRevLett.107.160501}, trapped-ions~\cite{Chiaverini:2004cj,Schindler2011,nigg_quantum_2014}, and superconducting qubits~\cite{Reed2012}. However these implementations lack the ability to perform fault-tolerant syndrome extraction which continues to be a challenge for all physical quantum computing systems.  Here we experimentally demonstrate a key step towards this problem by using a  two-by-two lattice of superconducting qubits to perform syndrome extraction and arbitrary error detection via simultaneous quantum non-demolition stabilizer measurements. This lattice represents a primitive tile for the surface code (SC)~\cite{Bravyi1998,Kit97-2}, which is a promising stabilizer code for scalable quantum computing. Furthermore, we successfully show the preservation of an entangled state in the presence of an arbitrary applied error through high-fidelity syndrome measurement. Our results bolster the promise of employing lattices of superconducting qubits for larger-scale fault-tolerant quantum computing.}

\begin{figure*}[t!]
\centering
\includegraphics[width=\textwidth]{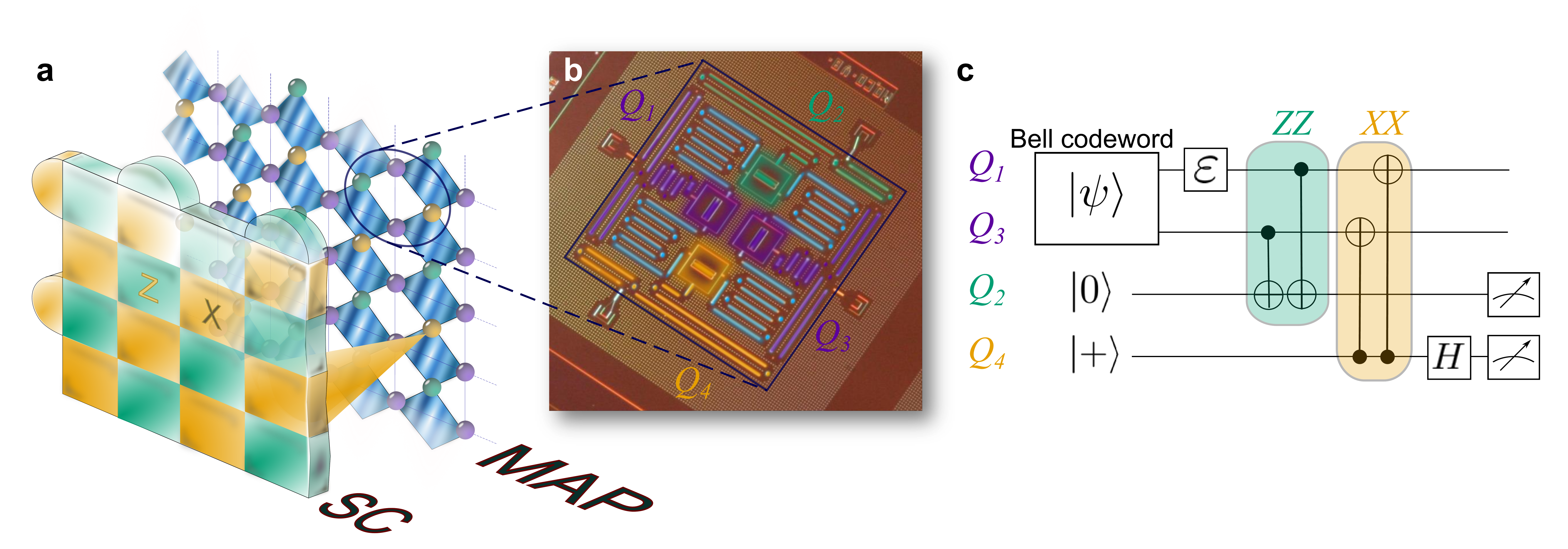}
\caption{\label{fig:1} \textbf{Surface code implementation and error detection quantum circuit.} \textbf{a}, Cartoon schematic of SC consisting of alternating square tiles of $X$- (yellow) and $Z$- (green) plaquettes for detecting phase-flip ($Z$) and bit-flip ($X$) errors, respectively. Semi-circular pieces reflect parity checks at the boundaries of the lattice. These plaquette tiles can be mapped onto a lattice of physical superconducting qubits with appropriate nearest-neighbour interconnectivity, as shown in the layer labeled MAP. Here, there are code qubits (purple spheres), $X$-syndrome qubits (yellow) for phase parity detection of surrounding code qubits, and $Z$-syndrome qubits (green) for bit parity detection of surrounding code qubits. The physical connectivity for superconducting qubits can be realised via coupling every qubit to two quantum bus resonators, shown as wavy blue diamonds in the MAP. The device studied in this work (false-colored optical micrograph in \textbf{b}) embodies two half-plaquettes of the SC as circled in \textbf{a}, and allows for independent and simultaneous detection of $X$ and $Z$ errors on two code qubits, shaded purple in \textbf{b} and labeled $Q_1$ and $Q_3$. \textbf{c}, The circuit to implement the half-plaquette operations encodes the bit ($ZZ$) and phase ($XX$) parities of the two code qubits' Bell state $|\psi\rangle$ onto the respective syndrome qubits, $Q_2$ (green) and $Q_4$ (yellow). Arbitrary errors $\varepsilon$ are intentionally introduced on the code qubit $Q_1$ and detected from the correlated measurement of the syndrome qubits. $Q_2$ ($Q_4$) is initialized to $\ket{0}$ ($\ket{+} = (\ket{0}+\ket{1})/2$). A Hadamard operation, $H$, is applied to $Q_4$ before measurement.}
\end{figure*}

The surface code (SC) has become a popular choice for physical implementation due to its nearest-neighbour qubit layout and high fault-tolerant error thresholds \cite{Raussendorf2007}. Code qubits are placed at the vertices of a two-dimensional array and each stabilizer involves four neighbouring code qubits (see Fig.~\ref{fig:1}a). These stabilizers are therefore geometrically local and can be measured fault tolerantly with a single syndrome qubit\cite{Dennis2002}. Error detection on a lattice of code qubits is achieved through mapping stabilizer operators onto a complementary lattice of syndrome qubits, followed by classical correlation of measured outcomes.  Amongst the syndrome qubits, a distinction is made between bit-flip syndromes (or $Z$-syndromes) and phase-flip syndromes (or $X$-syndromes). Each code qubit in the SC is coupled to two $X$-syndrome qubits and two $Z$-syndrome qubits, and in turn, each syndrome qubit is coupled to four code qubits.

Superconducting qubits have become prime candidates for SC implementation~\cite{fowler_surface_2012,chow_implementing_2014} especially with continuing improvements to coherence times~\cite{Paik2011,chang_improved_2013,Barends2013} and quantum gates~\cite{barends_superconducting_2014}. Furthermore, implementing superconducting resonators as quantum buses to realize the circuit quantum electrodynamics architecture permits a straightforward path to building connectivity into a lattice of superconducting qubits~\cite{chow_implementing_2014}. Although there are numerous ways of building the SC lattice with superconducting qubits and resonators, we employ an arrangement in which each qubit is coupled to two bus resonators, and each bus couples to four qubits~\cite{chow_implementing_2014}. The breakdown of the surface code into such an arrangement is shown in Fig.~\ref{fig:1}a. Although previously the engineered dissipation of a resonator has been used to stabilize the entanglement of two superconducting qubits to which it is coupled~\cite{shankar_autonomously_2013}, it is of note that here the stabilization is achieved via explicitly mapping code qubit stabilizers onto syndrome qubits.

For this work, we construct a four-qubit square lattice, which is a non-trivial cut-out of the SC layout (circled in Fig.~\ref{fig:1}a) and demonstrate both the $ZZ$ and $XX$ parity check. The $XX$ ($ZZ$) stabilizer is measured by the $X$-syndrome ($Z$-syndrome) qubit. Although previous work~\cite{Saira2014,chow_implementing_2014} implemented a $ZZ$ parity check on a linear arrangement of three qubits, our experiment goes beyond into the other planar dimension, allowing the demonstration of the $[[2,0,2]]$ code. Here, a codeword $\ket{\psi}=(\ket{00}+\ket{11})/\sqrt{2}=(\ket{++}+\ket{--})/\sqrt{2}$ is stored in two code qubits and protected from any single error to the codespace via syndrome detection. An arbitrary single-qubit error revealed in the stabilizer syndrome as a bit- (phase-) flip simply maps $\ket{\psi}$ to a negative eigenstate of $ZZ$ ($XX$) and a joint bit- and phase-flip ($Y$ rotation) maps $\ket{\psi}$ to the negative eigenstate of both $ZZ$ and $XX$. By encoding both the $XX$ and the $ZZ$ stabilizers in the four-qubit lattice, we can protect a maximally-entangled state of the two code qubits against an arbitrary error. 

Our physical device (Fig.~\ref{fig:1}b) consists of a $2\times2$ lattice of superconducting transmons, with each coupled to its two nearest neighbours via two independent superconducting coplanar waveguide (CPW) resonators serving as quantum buses (Fig.~\ref{fig:1}b, blue). Each qubit is further coupled to an independent CPW resonator for both qubit control and readout. Dispersive readout signals for each qubit are amplified by distinct Josephson parametric amplifiers (JPAs) giving high single-shot readout fidelity~\cite{johnson_heralded_2012,riste_initialization_2012}. We implement two-qubit echo cross-resonance gates~\cite{corcoles_process_2013}, $ECR = ZX_{90}-XI$, which are primitives for constructing controlled-NOT (CNOT) operations. Given the latticed structure of our device, we implement four different such gates, $ECR^{ij}$ between qubits $Q_i$ (control) and $Q_j$ (target), with $ij \in \{12,23,34,41\}$. In this arrangement, we use $Q_1$ and $Q_3$ (Fig.~\ref{fig:1}b, purple) as code qubits, $Q_2$ as the $Z-$syndrome qubit (Fig.~\ref{fig:1}b, green) and $Q_4$ as the $X-$syndrome qubit (Fig.~\ref{fig:1}b, yellow). All $ECR$ gates are benchmarked~\cite{corcoles_process_2013} with fidelities between ~0.93-0.97 (for further details see Methods). Single-qubit gates are benchmarked to fidelities above 0.998 with less than 0.001 reduction in fidelity due to crosstalk, as verified via  simultaneous randomized benchmarking~\cite{gambetta_characterization_2012}. The four independent single-shot readouts yield assignment fidelities (see Methods) all above $0.94$. These and other relevant system experimental parameters including qubit frequencies, anharmonicities, energy relaxation and coherence times are further discussed in the Methods.

To demonstrate the SC sub-lattice stabilizer measurement protocol (Fig.~\ref{fig:1}c), we first prepare the two code qubits in codeword state $\ket{\psi}$, which is a maximally entangled Bell state. Subsequently, the $ZZ$ stabilizer is encoded onto the $Z-$syndrome qubit $Q_2$, which is initialized in the ground state $\ket{0}$. The $XX$ stabilizer is encoded onto the $X-$syndrome qubit $Q_4$, which is initialized in the  $|+\rangle=(|0\rangle+|1\rangle)/\sqrt{2}$ state. Since we perform measurements of the syndrome qubits in the $Z$ measurement basis, $Q_4$ also undergoes a Hadamard transformation $H$ right before measurement. The complete circuit as shown in Fig.~\ref{fig:1}c will detect an arbitrary single-qubit error $\varepsilon$ to the code qubits via the projective measurements of the syndrome qubits.  We choose to apply the error on $Q_1$, but there is no loss of generality if applied on $Q_3$ instead. Each of the four possible outcomes of the syndrome qubit measurements projects the code qubits onto one of the four maximally entangled Bell states. If no error is present in the sequence, the syndrome qubits are both found to be in their ground state after the measurement, and the prepared codeword state of the code qubits is preserved. 

In our experiment, since the two code qubits ($Q_1$ and $Q_3$) are non-nearest neighbours in the lattice, the preparation of the codeword state is performed via two-qubit interactions with a shared neighbouring qubit, $Q_2$. The gate sequence for this state preparation can be compiled together with portions of the $ZZ$ stabilizer encoding. The resulting complete gate decomposition of the circuit from Fig.~\ref{fig:1}c in terms of our available single and two-qubit $ECR$ gates is described in detail in the Methods and shown in Extended Data Fig.~\ref{fig:supfigGates}.

\begin{figure}[t!]
\centering
\includegraphics[width=0.47\textwidth]{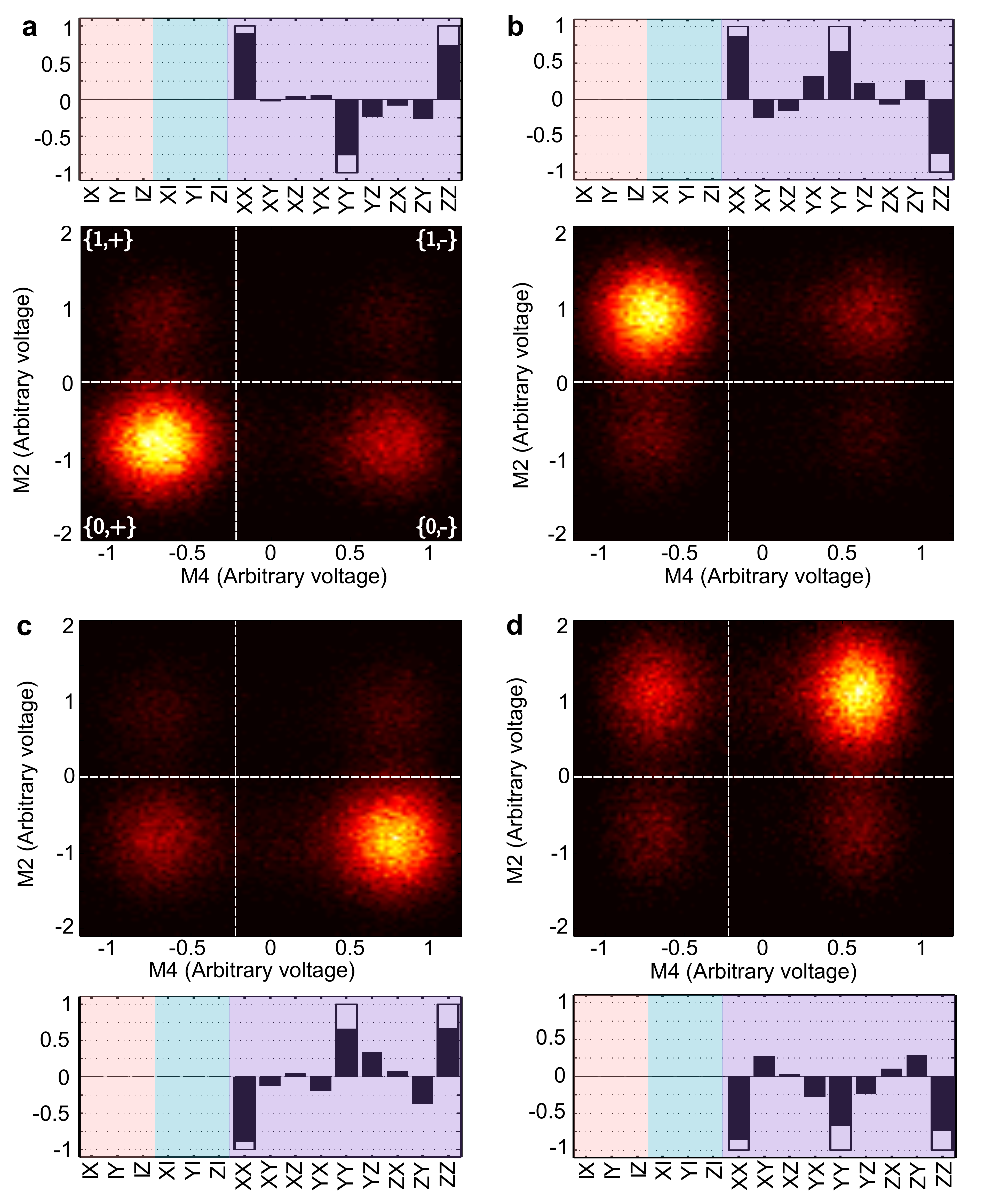}
\caption{\label{fig:2} \textbf{Correlated syndrome single-shot histograms and quantum state tomography of code qubits.} The quantum state of the syndrome qubits reveals the entangled state of the code qubits. The colormaps show the single-shot histograms of the syndrome measurements on $Q_2$ and $Q_4$. The dashed white lines indicate the threshold used to condition the reconstruction of the code qubit states, represented by a Pauli vector. The pink-, blue-, and purple-shaded regions signify $Q_1$, $Q_3$ and joint Pauli operators, respectively. Each of the possible four outcomes of correlated single-shot measurements of the syndrome qubits is mapped onto one of the four maximally entangled Bell states of the code qubits. Since we always prepare the code qubits in the codeword state $|\psi\rangle = (|00\rangle + |11\rangle)/\sqrt{2}$ at the beginning of the quantum process, when no error is applied to $Q_1$, state tomography of $Q_1$ and $Q_3$ conditioned on outcomes in the lower left quadrant $\{0,+\}$ of the colormap recover the same state with fidelity $0.8491\pm 0.0005$ (\textbf{a}). Introducing an error $\varepsilon$ equal to $X$ (b), $Z$ (c) and $Y$ (d) on $Q_1$, and conditioning on outcomes in the upper left $\{1,+\}$, lower right $\{0,-\}$, and upper right $\{1,-\}$ quadrants results in the code qubits reconstructed as $|\Psi\rangle = (|01\rangle + |10\rangle)/\sqrt{2}$ (fidelity $0.8195\pm 0.0006$), $|\Psi\rangle = (|00\rangle - |11\rangle)/\sqrt{2}$ (with fidelity $0.8046\pm 0.0005$), and $|\Psi\rangle = (|01\rangle - |10\rangle)/\sqrt{2}$ (fidelity $0.8148\pm 0.0006$), respectively. The $X-$syndrome qubit, $Q_4$, is found in its excited state when a phase-flip error has occurred (\textbf{c} and \textbf{d}) whereas the $Z-$syndrome qubit, $Q_2$, is found in its excited state as a result of bit-flip errors (\textbf{b} and \textbf{d}). The quoted uncertainties in reconstructed state fidelities are statistical (see Methods), but we note that systematic errors due to coherence time fluctuations, state preparation and measurement errors, can lead to indifelity $\sim0.01-0.02$.)}
\end{figure}

To implement arbitrary errors to the entangled code qubit state, we apply single-qubit rotations to $Q_1$ of the form $\varepsilon=U_{\theta}$, where $U$ defines the rotation axis and $\theta$ is the rotation angle (when no angle is given it is assumed $\theta = \pi$). Following the error detection protocol of Fig.~\ref{fig:1}c, we acquire single-shot measurements of the syndrome qubits and correlate  independent measurements around various axes of the code qubits for quantum state tomography~\cite{ryan_inprep2013}. First, for the case where $\varepsilon=U_{0}$, when no error is added, the two syndrome qubits $Q_2$ and $Q_4$ should both be measured to be in their ground states, and from correlating their single-shot measurements, $M_2$ and $M_4$, we detect the colour map as shown in Fig.~\ref{fig:2}a. Here, we can clearly see that a majority of the resulting measurements are located in the lower left quadrant, and we will use the notation $\{M_2,M_4\}=\{0,+\}$, with both syndromes signaling a ground state detection (note that measuring $Q_4$ in the ground state signals a $\ket{+}$ detection given the $H$ before measurement). Conditioned on $\{0,+\}$, state tomography of the code qubits is performed, with a reconstructed final state (Pauli vector shown in Fig.~\ref{fig:2}a), commensurate with the originally prepared codeword state with a fidelity of $0.8491\pm0.0005$. Next, for the case of a bit-flip error to $Q_1$, or $\varepsilon=X_{\pi}$, the resulting syndrome histograms are shown in the colour map Fig.~\ref{fig:2}b, where a majority of results are consistent with $\{1,+\}$, where the $Z$-syndrome $Q_2$ is excited to $\ket{1}$ and the $X$-syndrome $Q_4$  remains in its ground state. Conditioned on $\{1,+\}$, the reconstructed final state Pauli vector of the code qubits is now $(|01\rangle + |10\rangle)/\sqrt{2} =(|++\rangle - |--\rangle)/\sqrt{2} $, verifying the bit-flip parity error. Then, for the case of a phase-flip error on $Q_1$, or $\varepsilon=Z_{\pi}$, we find that the syndromes give $\{0,-\}$, with the $X$-syndrome having changed its state (Fig.~\ref{fig:2}c). Similarly, conditioned on  $\{0,-\}$, the code qubit state agrees with $(|00\rangle - |11\rangle)/\sqrt{2}=(|+-\rangle + |-+\rangle)/\sqrt{2}$, showing the phase-flip. Finally, an error $\varepsilon=Y_{\pi}$ results in both syndromes flipped, $\{1,-\}$, as shown in Fig.~\ref{fig:2}d with corresponding code qubit Pauli vector in agreement with both a bit- and phase-flip of the original codeword state $(|01\rangle - |10\rangle)/\sqrt{2}=(|+-\rangle - |-+\rangle)/\sqrt{2}$.

The reconstructed states reveal important information about our system. First, the measured state fidelity ($\sim 0.80$-$0.84$) is higher than expected ($\sim 0.75$)  from the measured fidelities of the five two-qubit gates and two independent single-shot measurements. This is because the gates used to prepare the codeword state do not contribute to the accumulated state fidelity loss, but rather reveal themselves as measurement errors. Secondly, the reconstructed conditional states have little to no weight in the single qubit subspace. This suggests that in our system there are negligible crosstalk errors (as expected since the code qubits are not directly connected via a bus). 

\begin{figure}[t!]
\centering
\includegraphics[width=0.47\textwidth]{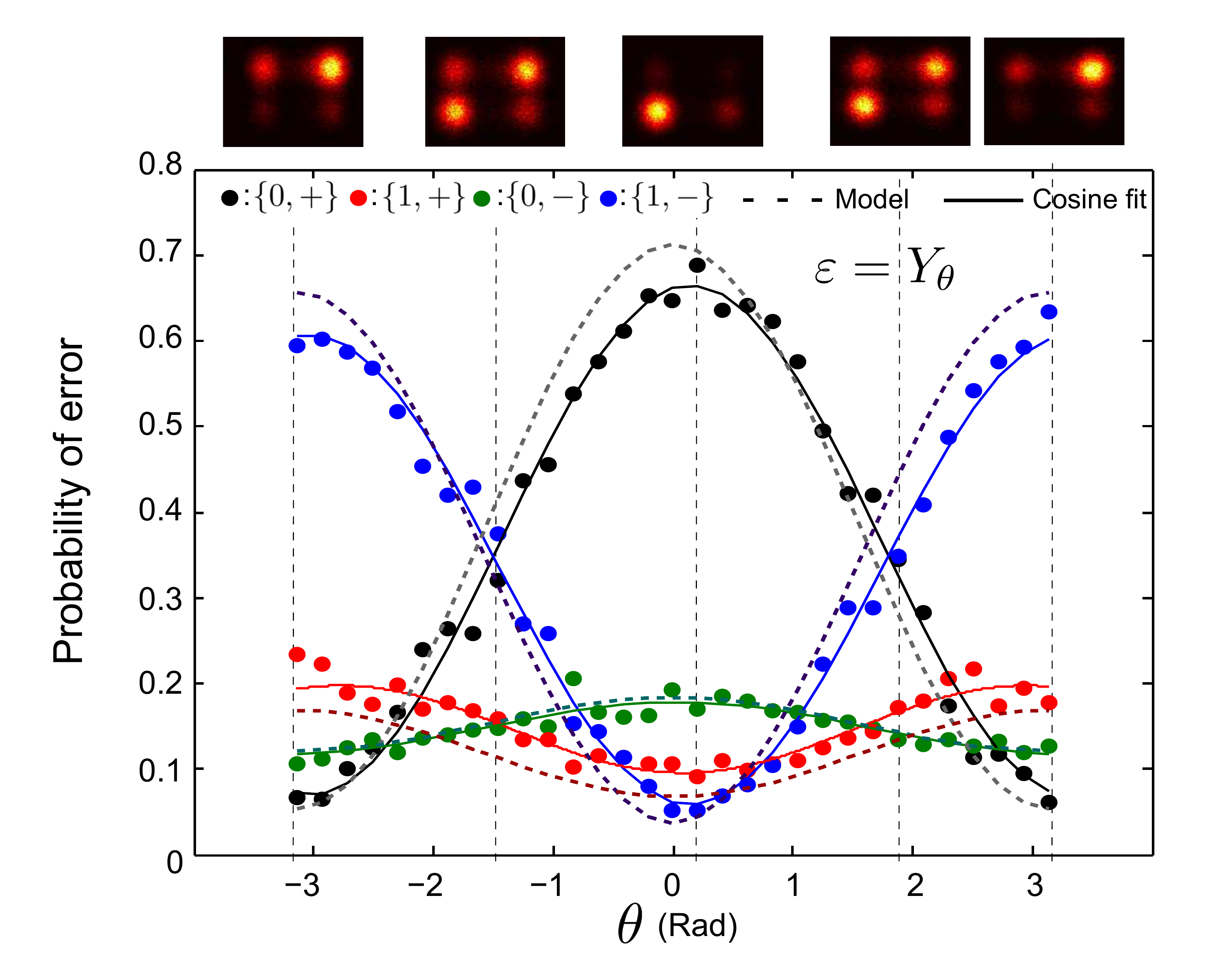}
\caption{\label{fig:3} \textbf{Syndrome qubits single-shot correlated measurement for different $Y-$error magnitudes.} The magnitude of each type of error in the code qubits can be extracted from the correlated single-shot traces of the syndrome qubits. Errors of $\varepsilon = Y_{\theta}$, with $\theta \in [-\pi,\pi]$ are detected by both syndromes, as $Y-$errors can be decomposed into a combination of bit- and phase-flip errors. As the magnitude of the $Y-$error increases from $0$ to $\pi$ the majority of the outcomes of the syndrome qubits changes from $\{M_2,M_4\}=\{0,+\}$ (black dots) to $\{M_2,M_4\}=\{1,-\}$ (blue dots) while the states $\{M_2,M_4\}=\{1,+\}$ (red dots) and $\{M_2,M_4\}=\{0,-\}$ (green dots), which indicate pure bit- and phase-flip errors, respectively, remain low probability. Solid lines are simple cosine fits to the data. Dashed lines are master-equation simulations that take into account the measured coherence times and assignment fidelities. Histograms of the correlated single-shot syndrome qubit measurements are shown in the density plots on top for $\theta \sim \{-\pi,-\pi/2,0,\pi/2,\pi\}$, as indicated by the vertical dashed-lines, with the syndrome states corresponding to $|\theta| = \pi/2$ showing significant populations in two quadrants.}
\end{figure}

We can track the outcome of the syndrome qubits as we slowly vary $\theta$ in an applied error $\varepsilon=Y_{\theta}$ between $-\pi$ and $+\pi$ (see Fig.~\ref{fig:3}). The state population of the four syndrome qubit states, $\{0,+\}$ (black dots), $\{1,+\}$ (red dots), $\{0,-\}$ (green dots) and $\{1,-\}$ (blue dots), obtained from the (normalised) number of counts in the correlated histograms conditioned on a readout threshold extracted from calibration measurements (see Methods), are plotted versus $\theta$. For an error induced by a unitary operation the data is explained by cosines (solid lines in Fig.~\ref{fig:3}). For $\theta$ near 0 the ground state, $\{0,+\}$, is found in both syndrome qubits as expected, whereas for $|\theta|\sim \pi$ we recover both syndrome flips $\{1,-\}$. The observed contrast between the different syndrome qubit state populations, near $0.6$ in Fig.~\ref{fig:3}, is commensurate with a master-equation simulation that takes into account the measured coherence times of our qubits and the assignment fidelities of the readouts (dashed lines in Fig.~\ref{fig:3}). Similarly, varying $\theta$ for $X$ and $Z$ rotations are shown in Extended Data Fig.~\ref{fig:SCosine}.

\begin{figure*}[t!]
\centering
\includegraphics[width=\textwidth]{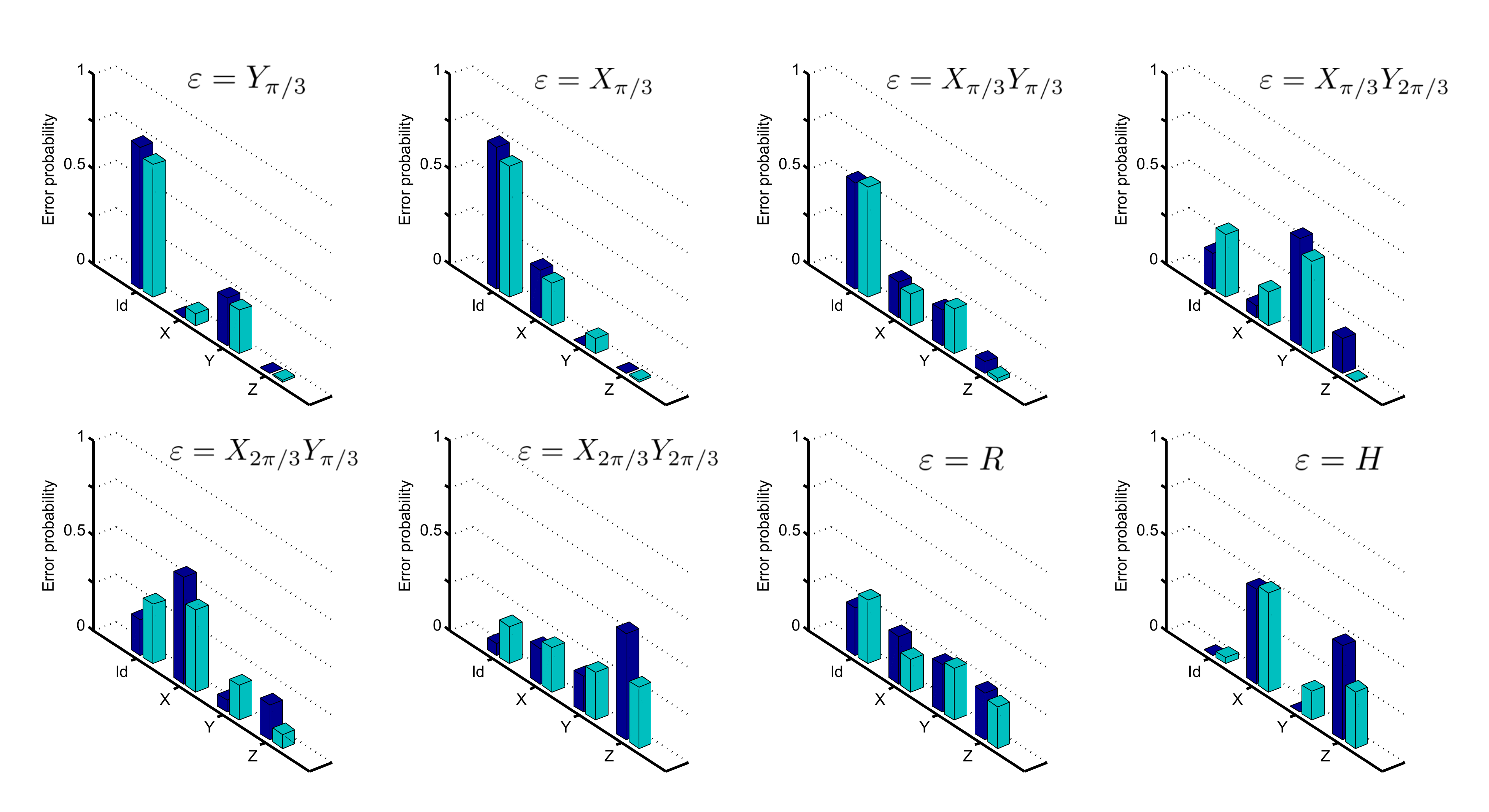}
\caption{\label{fig:4} \textbf{Detection of arbitrary errors.} The probability of each type of error, identity (Id), $X$, $Y$, or $Z$, is extracted from the correlated syndrome measurements for all the applied $\varepsilon$, as indicated above each panel. Dark blue bars represent the ideal outcome for each $\varepsilon$ and teal bars are measurements calibrated by the full $X$, $Y$, and $Z$ error rotation curves. The errors labeled as $R$ and $H$ correspond to a $Y_{\pi/2}X_{\pi/2}$ operation, which maps the $x-y-z$ axes in the Bloch sphere to $y-z-x$, and the Hadamard gate, respectively. The results are consistent with a higher uncertainty in the phase-flip error detection, likely due to decoherence during the full sequence and the order of syndrome detection.}
\end{figure*}

In order to demonstrate arbitrary error detection, we construct $\varepsilon$ via combinations of $X$ and $Y$ errors. Each panel of Fig.~\ref{fig:4} shows a teal bar plot reflecting the experimentally extracted population of each of the four possible syndrome qubit measurement outcomes for the set of errors $\{Y_{\pi/3}$, $X_{\pi/3}$, $X_{\pi/3}Y_{\pi/3}$, $X_{\pi/3}Y_{2\pi/3}$, $X_{2\pi/3}Y_{\pi/3}$, $X_{2\pi/3}Y_{2\pi/3}$, $R$, $H\}$, where $R=Y_{\pi/2}X_{\pi/2}$ and $H$ is the Hadamard operation. Overall, we find decent agreement between the experiment and ideal population outcomes (dark blue bars). The measured populations are renormalised by the observed contrast at $\theta\sim 0$ in Fig.~\ref{fig:3}, and in the equivalent plots for $X$- and $Z$-errors in Extended Data Fig.~\ref{fig:SCosine}, to account for relaxation and decoherence fidelity loss. Although this renormalisation provides an overall fairer comparison to the ideal case, it tends to increase the uncertainty in the bars, especially for the $Z-$error due to the loss of contrast and larger data scatter observed for the $X-$syndrome qubit $Q_4$ (see Methods, Extended Data Fig.~\ref{fig:SCosine}b). This diminished contrast is due to the fact that the $XX$ stabiliser is encoded last in our sequence and therefore suffers more from decoherence. Further improvements to coherence times and assignment fidelities will translate to more accurate error detection.

In conclusion, we have experimentally realised arbitrary quantum error detection using a two-by-two lattice of superconducting qubits. Stabilizer measurements, ubiquitous to fault-tolerant quantum error correcting codes, are successfully demonstrated for both bit- and phase-flip errors on an encoded codeword. The error detection experiments presented constitute a key milestone for surface code implementation, as our operations now extend into the plane of the two-dimensional surface and we show the ability to concurrently perform bit- and phase- parity checks. Moreover, our results illustrate the ability to build structures of superconducting qubits which are not co-linear but latticed while preserving high-fidelity operations. Moving forward, further expanding the lattice will lead to important studies of different error correcting codes and the encoding of logical qubits, thereby allowing experimental investigation of fault-tolerant quantum computing.

\section*{Methods Summary}

\subsection*{Device fabrication}
The device is fabricated on a 720\,$\mu$m thick Si substrate. The superconducting CPW resonators, the qubit capacitors, and coupling capacitors are defined in the same step via optical lithography. Reactive ion etching of a sputtered 200 nm thick Nb film is used to make this layer. The Josephson junctions, patterned via electron beam lithography, are made by double-angle deposition of Al (layer thicknesses of 35 and 85 nm) followed by a liftoff process. The chip is mounted on a printed circuit board and wirebonded for signal delivery and cross-talk mitigation.

\subsection*{Device parameters}
The four qubit transition frequencies are $\omega_i/2\pi=\{5.303, 5.101, 5.291, 5.415\}$ GHz with $i\in\{1,2,3,4\}$. The readout resonator frequencies are $\omega_{Ri}/2\pi=\{6.494, 6.695, 6.491, 6.693\}$ GHz whilst the four bus resonators, unmeasured, are designed to be at $\omega_{Bij}/2\pi=\{8,7.5,8,7.5\}$ GHz for $ij\in\{12,23,34,41\}$. All qubits show around 330 MHz anharmonicity, with energy relaxation times $T_{1 (i)} = \{33, 36, 31, 29\}$ $\mu$s and coherence times $T_{2 (i)}^{\rm{echo}} = \{17, 16,18,22\}$ $\mu$s. The dispersive shifts and line widths of the readout resonators are measured to be $2\chi_i/2\pi = \{-3.0,-2.0,-2.5,-2.8\}$ MHz and $\kappa_i/2\pi = \{615,440,287,1210\}$ kHz, respectively. 

\section*{Methods}

\subsection*{Gate calibration and characterisation}

Single qubit gates are 53.3 ns long Gaussian pulses with width $\sigma =13.3$ ns. We use single-sideband modulation to avoid mixer leakage at the qubit frequencies in between operations. The sideband frequencies, which are chosen taking into account all qubit frequencies and anharmonicities, are +60, -80, +180 and +100 MHz for $Q_1$, $Q_2$, $Q_3$ and $Q_4$, respectively. Every single-qubit pulse is accompanied by a scaled Gaussian derivative in the other quadrature to minimise the effect of leakage of information into higher qubit energy levels~\cite{Motzoi:2009fx}. All microwave mixers are independently calibrated at the operational frequencies to minimise carrier leakage as well as to ensure orthogonality of the quadratures. Following these calibrations, the single-qubit rotations are tuned by a series of repeated rotations described elsewhere~\cite{Chow2012PRL}.

Randomised Benchmarking (RB) of single-qubit gates~\cite{Magesan2011} is performed for all four qubits independently and in all possible simultaneous configurations (Table~\ref{table:S1}). This allows us to establish the degree of addressability error~\cite{gambetta_characterization_2012} present in our system. Comparing the individual and simultaneous RB experiments we can see that the addressability error is $0.001$ or lower in all cases.

\begin{suppfigure}[htbp!]
\centering
\includegraphics[width=0.47\textwidth]{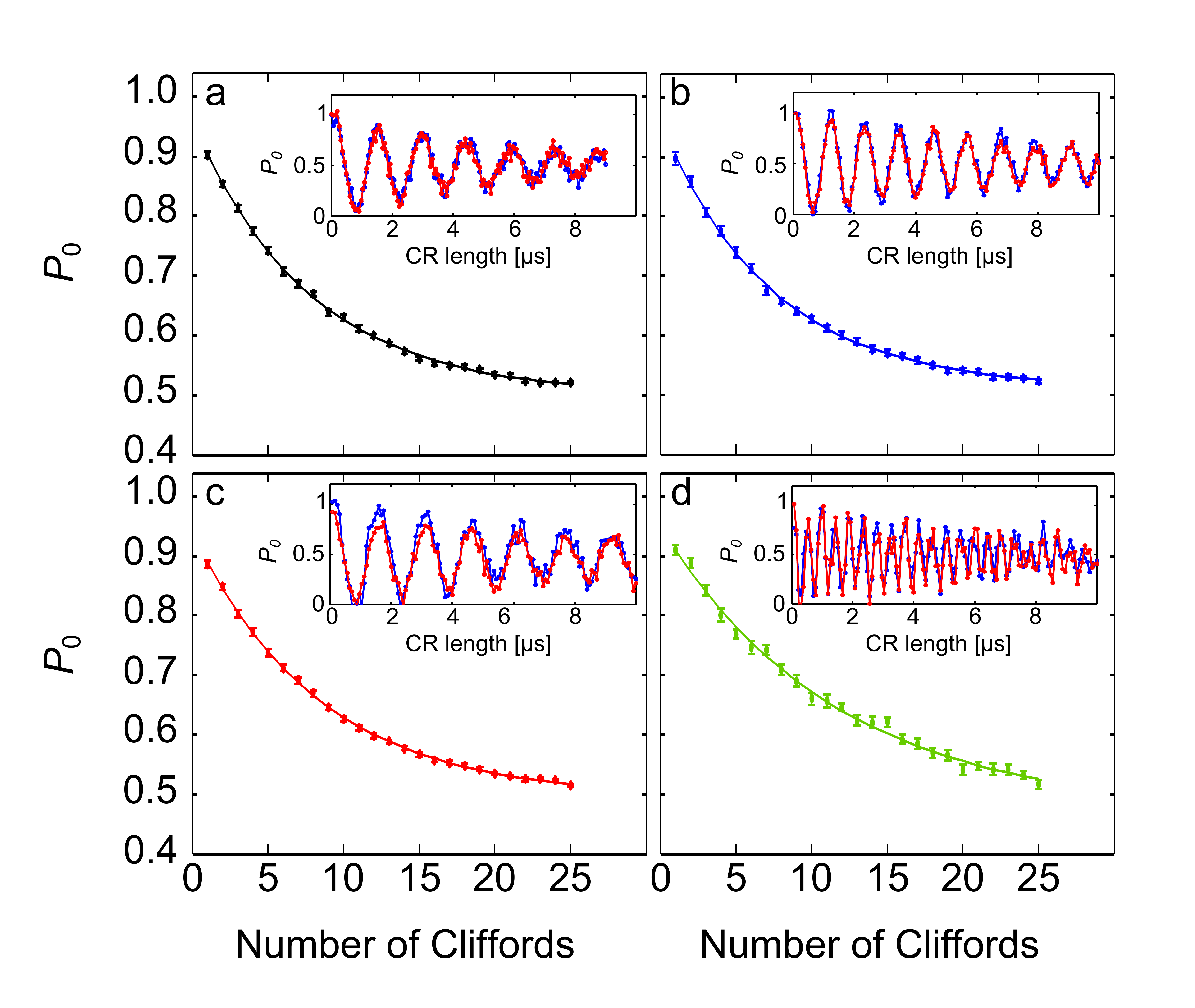}
\caption{\label{fig:S2} \textbf{Two-qubit randomized benchmarking.} Average population of the ground state of the target qubit, $P_0$, versus number of two-qubit Cliffords generated via $ECR$ gates between (a)~$Q_1$ and $Q_2$, (b)~$Q_2$ and $Q_3$, (c)~$Q_3$ and $Q_4$, and (d)~$Q_4$ and $Q_1$. Each RB experiment is averaged over 50 different sequences. Fits to the experiments are shown as solid lines and yield average errors per two-qubit Clifford of (a)~$0.0604\pm0.0006$, (b)~$0.0631\pm 0.0007$, (c)~$0.0569\pm 0.0015$, and (d)~$0.0353\pm 0.0015$. Inset shows $ZX$ oscillations~\cite{corcoles_process_2013} of the target qubit state population as a function of the cross-resonance drive length when the control qubit is in the ground (blue) or in the excited (red) state.}
\end{suppfigure}

\begin{table}
\begin{ruledtabular}
\begin{tabular}{|c|c c c c|}
Qubit label & $M_1$ [$\times 10^{-3}$] & $M_2$ [$\times 10^{-3}$] & $M_3$  [$\times 10^{-3}$] & $M_4$  [$\times 10^{-3}$]\\ 
\hline
0001 	& --  	& -- 	& -- & $1.07\pm0.03$\\
0010 	& --		& --  & $1.18\pm0.03$ & --\\ 
0100	& --		& $1.22\pm 0.03$	& -- & --\\
1000	& $1.37\pm 0.03$ &	-- &	-- & --\\

0011 	& -- & -- & $1.47\pm0.03$ & $1.15\pm0.03$\\
0101 	& -- & $1.33\pm0.04$ & -- & $1.16\pm0.02$\\ 
1001 	& $1.35\pm0.03$ & -- & -- & $1.10\pm0.03$\\

0110	& -- & $1.40\pm0.04$ & $1.58\pm0.03$ & --\\
1010 	& $1.73\pm0.05$ & -- & $2.06\pm0.05$ & --\\
1100 	& $1.30\pm0.05$ & $1.35\pm0.03$ & -- & --\\ 

0111	& -- & $1.49\pm0.03$ & $1.45\pm0.02$ & $1.18\pm0.03$\\
1011 	& $1.90\pm0.07$ & -- & $2.22\pm0.08$ & $1.12\pm0.03$\\
1101 	& $1.36\pm0.03$ & $1.41\pm0.04$ & -- & $1.06\pm0.04$\\ 
1110	& $1.90\pm0.05$ & $1.32\pm0.03$ & $1.98\pm0.06$ & --\\
1111	& $2.07\pm0.05$ & $1.37\pm0.05$ & $2.12\pm0.06$ & $1.17\pm0.04$\\
\end{tabular}
\end{ruledtabular}
\caption{\label{table:S1} \textbf{Summary of simultaneous single-qubit RB.} The ones in the first column indicate which qubits in the string $Q_1Q_2Q_3Q_4$ are being randomised and the zeros which qubits are being left unperturbed.}
\end{table}

The two-qubit $ECR$ gates consist of two cross-resonance pulses of different signs, each of duration $\tau$, separated by a $\pi$ rotation in the control qubit. This sequence selectively removes the $IX$ part of the Hamiltonian whilst enhancing the $ZX$ term~\cite{corcoles_process_2013}. Each cross-resonance pulse has a Gaussian turn-on and off of width 3$\sigma$ with $\sigma = 24$ ns, included in $\tau$. The gates $ECR^{12}$, $ECR^{23}$, $ECR^{34}$, and $ECR^{41}$, where $ECR^{ij}$ is the $ECR$ gate between $Q_i$ (control) and $Q_j$ (target), had $\tau$ of 400, 360, 440 and 190 ns, respectively, for a total gate time of $2\times \tau + 53.3$ ns. We also characterise the two-qubit gates via Clifford RB~\cite{corcoles_process_2013}. Extended Data Fig.~\ref{fig:S2} shows the RB decays for each of the four gates, yielding an error per two-qubit Clifford gate of $0.0604\pm0.0006$, $0.0631\pm 0.0007$, $0.0569\pm 0.0015$ and $0.0353\pm 0.0015$ for $ECR^{12}$, $ECR^{23}$, $ECR^{34}$, and $ECR^{41}$, respectively.

\begin{suppfigure*}[htbp!]
\centering
\includegraphics[width=\textwidth]{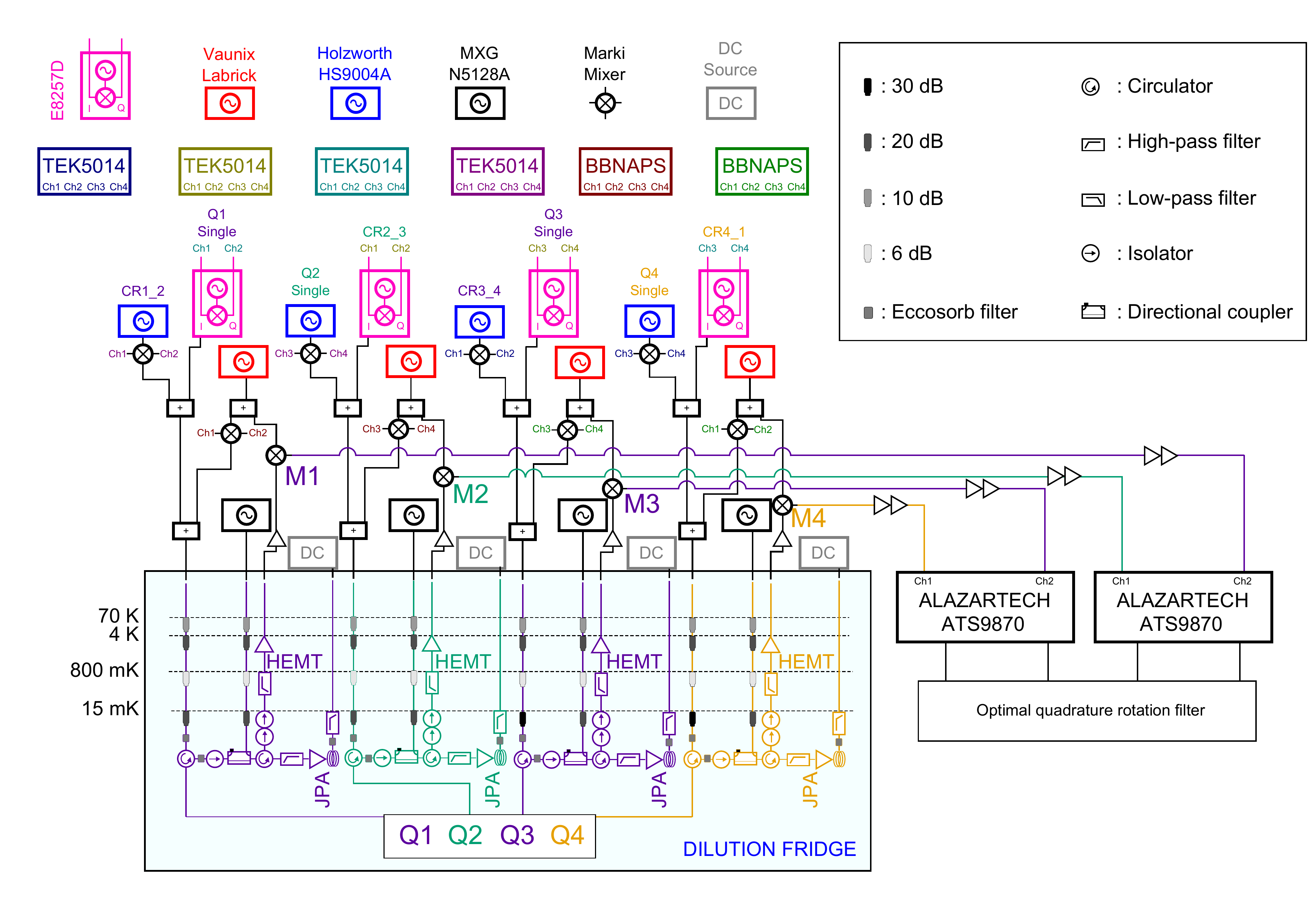}
\caption{\label{fig:S1} \textbf{Experimental setup.} Detailed wiring scheme for all room temperature control electronics and internal configuration of the Oxford Instruments Triton dilution refrigerator.}
\end{suppfigure*}

\subsection*{Experimental setup}

We cool our device to 15 mK in an Oxford Triton dilution refrigerator. Extended Data Fig.~\ref{fig:S1} shows a full schematic of the measurement setup. We achieve independent single-shot readout for each qubit by using a HEMT amplifier following a JPA (provided by UC Berkeley) in each readout line. The device is protected from environmental radiation by an Ammuneal cryoperm shield with an inner coat of Emerson \& Cuming CR-124 eccosorb. All qubit control lines are heavily attenuated at different thermal stages and home-made eccosorb microwave filters are added at the coldest refrigerator plate.

Single-qubit and two-qubit control pulses as well as resonator readout pulses are generated using single-sideband modulation. The modulating tones are produced by Tektronix arbitrary waveform generators (model AWG5014) for qubit operations. Modulating shapes for readout are produced by Arbitrary Pulse Sequencers from Raytheon BBN Technologies. We either use external Marki I/Q mixers with a Holzworth microwave generator, or an Agilent vector signal generator (E8257D) as depicted in Extended Data Fig.~\ref{fig:S1}. For data acquisition we use two AlazarTech two-channel digitizers (ATS9870) and the single-shot readout time traces are processed with an optimal quadrature rotation filter~\cite{ryan_inprep2013}.

\subsection*{Circuit gate decomposition}

\begin{suppfigure*}[htbp!]
\centering
\includegraphics[width=\textwidth]{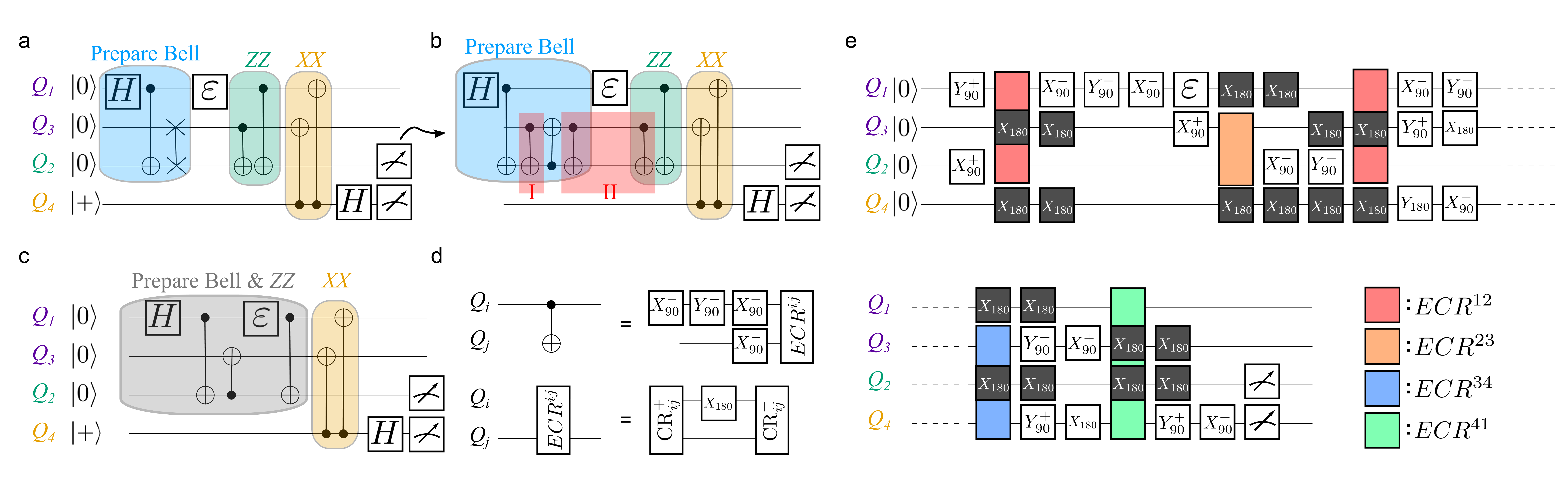}
\caption{\label{fig:supfigGates} \textbf{Quantum circuit and CNOT gate decomposition.} The $ZZ$ and $XX$ parity checks are performed on a pair of maximally entangled qubits. This entanglement is achieved in our architecture with one CNOT and one SWAP gate (a). The three CNOTs that define the SWAP gate can be combined with the following $ZZ$ parity check operation to simplify the circuit and three CNOT gates can be eliminated (b). The final circuit implemented in our experiments has a total of five CNOT gates (c). We implement our CNOT gates by using a simplified version of the $ZX_{90}^{ij}$ gate, $ECR^{ij}$, consisting of two cross-resonance pulses of different sign separated by a $\pi$ rotation in the control qubit. With that definition, a CNOT gate can be obtained with four single-qubit rotations plus a $ECR^{ij}$ operation. An example, not unique, of such decomposition is shown in (d). The complete gate sequence in our error detection experiments is presented in (e), where the dark boxes indicate refocus pulses during every two-qubit gate on the two qubits not involved on it.}
\end{suppfigure*}

The circuit in Fig.~\ref{fig:1}c calls for four two-qubit gates. In addition, the code qubits $Q_1$ and $Q_3$ need to be prepared in an entangled state. Since these qubits are not nearest neighbours and there is no provision for interaction between them -a key feature of the SC-, we first entangle $Q_1$ and $Q_2$ and then perform a swap operation between $Q_2$ and $Q_3$ (Extended Data Fig.~\ref{fig:supfigGates}a). A SWAP gate operation is equivalent to three CNOT gates alternating direction (Extended Data Fig.~\ref{fig:supfigGates}b). Since two consecutive identical CNOT gates are equal to the identity operation and $Q_3$ starts from the ground state, the red shadowed regions in Extended Data Fig.~\ref{fig:supfigGates} can be omitted. The actual circuit implemented in our experiments is shown in Extended Data Fig.~\ref{fig:supfigGates}c, where the Bell state preparation and the $ZZ$ encoding have been combined. 

Our CNOT operations require an entangling gate between the control and the target qubits. We use the $ECR^{ij}$ as our CNOT genesis. The $ECR^{ij}$ gate plus four single-qubit rotations as depicted in Extended Data Fig.~\ref{fig:supfigGates}c correspond to a CNOT operation between $Q_i$ and $Q_j$ in our device.

\begin{suppfigure*}[htbp!]
\centering
\includegraphics[width=\textwidth]{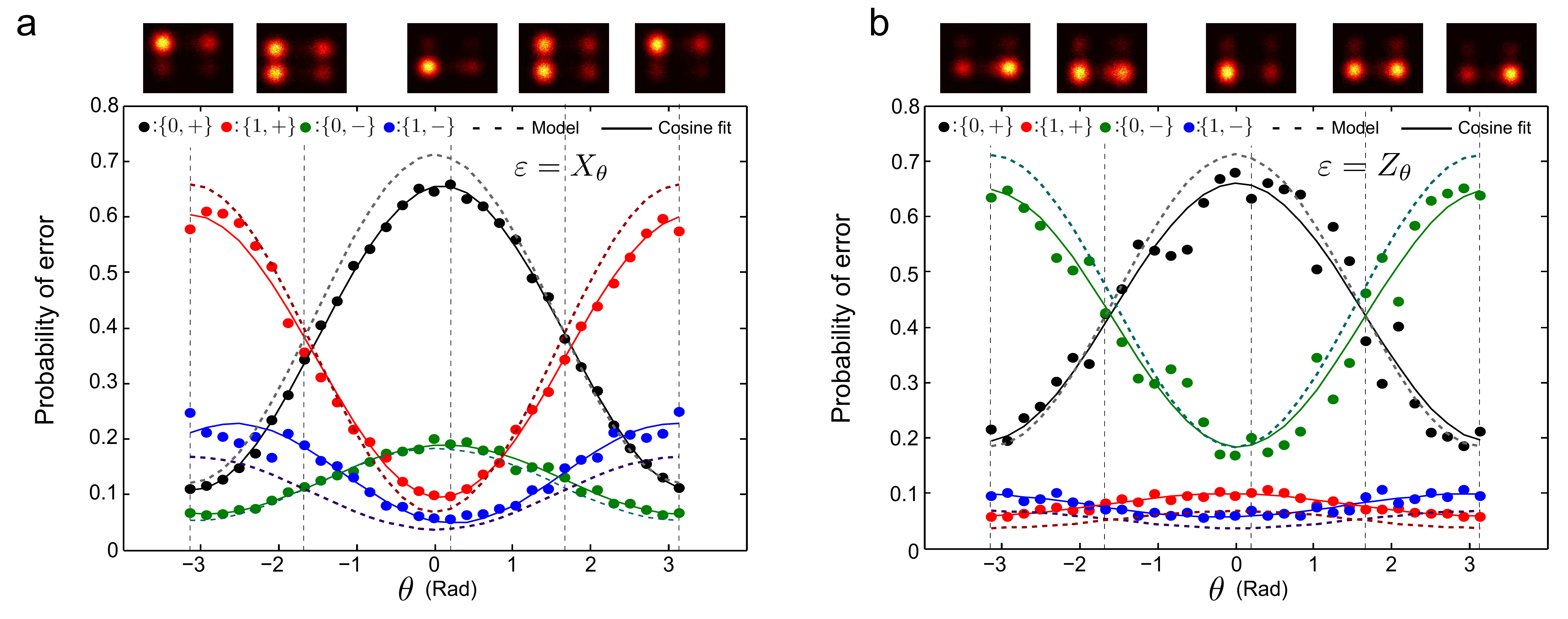}
\caption{\label{fig:SCosine} \textbf{Continuous tracking of pure bit- and phase-flip errors.} Errors of $\varepsilon = X_{\theta}$ (a) and $\varepsilon = Z_{\theta}$ (b) with $\theta \in [ -\pi, \pi ]$ are applied to the code qubit $Q_1$. The syndrome qubit states $\{M_2,M_4\}=\{0,+\}$ (black), $\{M_2,M_4\}=\{0,-\}$ (green), $\{M_2,M_4\}=\{1,+\}$ (red) and $\{M_2,M_4\}=\{1,-\}$ (blue) indicate the magnitude and nature of the error $\varepsilon$. Since the $ZZ$ and $XX$ parities are encoded into $Q_2$ and $Q_4$, respectively, pure bit-flip errors are detected by $Q_2$ whereas pure phase-flip errors are detected by $Q_4$.}
\end{suppfigure*}

\subsection*{Tracking bit- and phase-flip errors}

As introduced in the main text (Fig.~\ref{fig:3}), we can measure the magnitude of the error $\varepsilon$ from the correlated single-shot traces of the syndrome qubits. Here we show the figures complementing Fig.~\ref{fig:3} in the main text, corresponding to pure bit-flip error (Extended Data Fig.~\ref{fig:SCosine}a) and pure phase-flip error (Extended Data Fig.~\ref{fig:SCosine}b). We attribute the increased loss of contrast in the phase-flip error detection (Extended Data Fig.~\ref{fig:SCosine}b) to the order of the stabilizer encoding in our circuit, which makes our error detection protocol less sensitive to phase-flip errors.

\subsection*{Error propagation and syndromes}

After the SWAP gate and error in the circuit in Extended Data Fig.~\ref{fig:supfigGates}a, the state of the code qubits is given by $\varepsilon(|00\rangle + |11\rangle)/\sqrt{2}$ where $\varepsilon$ is some unitary operator acting on the first code qubit. We will find how the different Pauli errors propagate through the rest of the circuit to produce the different error syndromes.

First, suppose $\varepsilon$ is the bit-flip operation $X_{C_1}$ on the first code qubit $C_1$. In this case 
\begin{align}
X_{C_1} &\rightarrow \textrm{CNOT}_{C_1,S_1}X_{C_1}\textrm{CNOT}_{C_1,S_1}\nonumber \\
&= X_{C_1} X_{S_1} \nonumber \\
 &\rightarrow H_{S_2}\textrm{CNOT}_{S_2,C_1}X_{C_1}\textrm{CNOT}_{S_2,C_1}H_{S_2}\nonumber \\
&= X_{C_1} X_{S_1}.
\end{align}
where the sub-indexes $S_1$ and $S_2$ refer to the $Z-$ and $X-$syndrome qubits, $Q_2$ and $Q_4$ in our experiment, respectively. Similarly for the phase-flip operation $Z_{C_1}$,
\begin{align}
Z_{C_1} &\rightarrow \textrm{CNOT}_{C_1,S_1}Z_{C_1}\textrm{CNOT}_{C_1,S_1}\nonumber \\
&= Z_{C_1} \nonumber \\
 &\rightarrow H_{S_2} \textrm{CNOT}_{S_2,C_1}Z_{C_1}\textrm{CNOT}_{S_2,C_1} H_{S_2} \nonumber \\
&= Z_{C_1} X_{S_2},
\end{align}
and for $Y_{C_1}$
\begin{align}
Y_{C_1} &= -iZ_{C_1}X_{C_1}\nonumber \\
 &\rightarrow -i\textrm{CNOT}_{C_1,S_1}Z_{C_1}X_{C_1}\textrm{CNOT}_{C_1,S_1}\nonumber \\
 &= -iZ_{C_1} \textrm{CNOT}_{C_1,S_1}X_{C_1}\textrm{CNOT}_{C_1,S_1}\nonumber \\
&=-i Z_{C_1}X_{C_1}X_{S_1} \nonumber \\
 &\rightarrow -i H_{S_2} \textrm{CNOT}_{S_2,C_1}Z_{C_1}X_{C_1}X_{S_1}\textrm{CNOT}_{S_2,C_1} H_{S_2} \nonumber \\
& =  -i H_{S_2} \textrm{CNOT}_{S_2,C_1}Z_{C_1}\textrm{CNOT}_{S_2,C_1} X_{C_1}X_{S_1} H_{S_2} \nonumber \\
&=  -i H_{S_2} Z_{C_1}Z_{S_2} X_{C_1}X_{S_1} H_{S_2}\nonumber \\
&= -i Z_{C_1}X_{S_2} X_{C_1}X_{S_1} \nonumber \\
&= Y_{C_1}X_{S_1}X_{S_2}.
\end{align}

Since the state after the SWAP gate is $ |00;00\rangle + |11;00\rangle$ (the qubits are ordered $|Q_1Q_3;Q_2Q_4\rangle$ and $Q_1$, $Q_3$ are the code qubits) the error syndromes are given by
\begin{align}
\text{No error} &: |00;00\rangle + |11;00\rangle \rightarrow 00, \nonumber \\
\text{X error} &: |01;10\rangle + |10;10\rangle \rightarrow 10,\nonumber \\
\text{Z error} &: |00;01\rangle - |11;01\rangle \rightarrow 01,\nonumber \\
\text{Y error} &: |01;11\rangle - |10;11\rangle \rightarrow 11.
\end{align}
Hence if the error is a general single-qubit unitary operation
\begin{align}
U=\exp\left(-i\frac{\theta}{2}\hat{n}\cdot \vec{\sigma}\right),
\end{align}
the different error syndromes have the following probabilities of occurring
\begin{align}
 \text{No error} &: \text{pr}(00) =  \cos ^2\left(\frac{\theta}{2}\right), \nonumber \\
\text{X error} &: \text{pr}(10) = \sin ^2\left(\frac{\theta}{2}\right) n_x^2, \nonumber \\
\text{Z error} &: \text{pr}(01)  = \sin ^2\left(\frac{\theta}{2}\right) n_y^2, \nonumber \\
\text{Y error} &: \text{pr}(11) = \sin ^2\left(\frac{\theta}{2}\right) n_z^2, \label{eq:error_probs}
\end{align} where $n_i$ is the $i^\mathrm{th}$ component of the unit vector.

\subsection*{Readout characterization}

To characterize each readout, we create the $2^4=16$ standard computational basis (calibration) states and record the full time-dependent trajectory of the state of the cavity over a measurement integration time of $3 \mu$s. This process is repeated $19200$ times to gather sufficient statistics. Integrating kernels are obtained for each measurement channel which extract the full time-dependent readout information~\cite{ryan_inprep2013}. Histograms are fitted to the integrated shots and thresholds for each channel are set at the point of maximum distance between cumulative distributions of the histograms.

The assignment fidelity of each channel is calculated according to the standard formula
\begin{equation}
\mathcal{F}_{a} = 1 -P(0|1)/2-P(1|0)/2,
\end{equation} 
where $P(0|1)$ ($P(1|0)$) is the probability of obtaining ``0" (``1") when state $|1\rangle$ ($|0\rangle$) is created. The assignment fidelities are given in Table~(\ref{table:Assfids}).

\begin{table}[ht]
\caption{Assignment fidelities of the readout channels} 
\centering 
\begin{tabular}{ c c c c} 
\hline\hline 
 $M_1$ & $M_2$ & $M_3$ & $M_4$ \\ [0.5ex] 

\hline 
 0.9592 & 0.9476 & 0.9416 & 0.9646 \\ [1ex]
\hline 
\end{tabular}
\label{table:Assfids}
\end{table}

\subsection*{State tomography}

The conditional states of the code qubits ($Q_1$ and $Q_3$) for the different error types ($I$, $X$, $Y$, and $Z$) were reconstructed by applying the complete set of 36 unitary rotations $\mathcal{U}_M=\{\mathcal{I},X_{90},X_{\pm 45},Y_{\pm 45} \}^{\otimes 2}$ to the state of code qubits to attain a complete set of measurement operators. The fundamental measurement observables $\mathcal{O}_1,\mathcal{O}_2,\mathcal{O}_3$ that are rotated by elements of $\mathcal{U}_M$ are constructed from the calibration states by first normalizing the shots for each of the code qubit channels to lie in $[-1,1]$ and then correlating the shots. Note that if the calibrations were perfect then $\mathcal{O}_1,\mathcal{O}_2,\mathcal{O}_3$ are equal to $ZI$, $IZ$, and $ZZ$.

For each of the 36 different measurement settings, we bin each shot according to the measurement results of the syndrome qubits $Q_2$ and $Q_4$. As there are two syndrome qubits, there are four bins labeled by down-down, down-up,up-down,up-up. Denoting the conditional states $\rho^{dd}$, $\rho^{du}$, $\rho^{ud}$, and $\rho^{uu}$ we have full tomographic information of the state of the code qubits for each of the four bins. The shots are correlated to create the expectation values of each conditional state. Hence for each $U \in \mathcal{U}_M$, label $ab$, and observable $\mathcal{O}_j$ we have an estimate of $\text{trace}(U\rho^{ab}U^{\dagger}\mathcal{O}_j)$.

For each label $ab$ we have a measurement vector $m^{ab}$ of length 108 (36 unitary rotations $\times$ 3 fundamental observables). Choosing any representation $x^{ab}$ of $\rho^{ab}$ in some operator basis allows us to write
\begin{align}
m^{ab}&= M x^{ab}
\end{align}
where $M$ is a constant matrix whose entries depend only on the choice of operator basis. We choose to use the standard Pauli basis to represent $\rho^{ab}$,
\begin{align}
\rho^{ab}&=\sum_{j=0}^{15}x^{ab}_jP_j,
\end{align}
which implies $M$ is a $108\times 16$ matrix. Enforcing $\rho^{ab}$ to be trace 1 sets $x_0=1/4$.

$x^{ab}$ can be solved for in a variety of ways, the most straightforward of which is linear inversion via computing the pseudo-inverse of $M$. While linear inversion provides a valid statistical estimator, it does not enforce positivity of the state. Alternatively, one can maximize the likelihood function for the measurement results under the assumption of Gaussian noise~\cite{Chow2012PRL} and solve the following constrained quadratic optimization problem
\begin{align}
&\text{argmin}_x  \: \: \left\|\sqrt{{\left(V^{ab}\right)}^{-1}}\left(m^{ab}-Mx\right)\right\|_2^2 \nonumber \\
&\text{subject to:} \: \:  I/4 + \sum_{j=1}^{15} x_j P_j \geq 0,
\end{align}
to obtain a physically valid state. Here $V^{ab}$ is the variance matrix of the measurement matrix. When only Gaussian noise is present, solving this optimization problem is equivalent to finding the closest physical state to the linear inversion estimate~\cite{Smolin2012}.

We quantify the state reconstruction via the state fidelity between $\rho_\mathrm{noisy}=\rho^{ab}$ and the ideal target state $\rho_\mathrm{ideal}=|\psi^{ab}\rangle\langle \psi^{ab} |$;
\begin{equation}
\mathcal{F}_\mathrm{state} (\rho_\mathrm{noisy},\rho_\mathrm{ideal}) = \left(\mathrm{Tr}[\sqrt{\sqrt{\rho_\mathrm{ideal  }}{\rho_\mathrm{noisy  }}\sqrt{\rho_\mathrm{ideal  }}}]\right)^2,
\end{equation} 
where, as mentioned in the main text, the ideal states for the different syndrome results are given by
\begin{align}
&00 \: (\text{no error}): |\psi^{00}\rangle = \frac{1}{\sqrt{2}}\left(|00\rangle + |11\rangle\right), \nonumber \\
&01 \: (\text{Z error}): |\psi^{01}\rangle = \frac{1}{\sqrt{2}}\left(|00\rangle - |11\rangle\right), \nonumber \\
&10 \: (\text{X error}): |\psi^{10}\rangle = \frac{1}{\sqrt{2}}\left(|01\rangle + |10\rangle\right), \nonumber \\
&11 \: (\text{Y error}): |\psi^{11}\rangle = \frac{1}{\sqrt{2}}\left(|01\rangle - |10\rangle\right).
\end{align}
The results are contained in Table~\ref{table:Fid}. The variance in the state fidelity is computed via a bootstrapping protocol described in~\cite{Chow2012PRL} and the physicality is the sum of the negative eigenvalues of the linear inversion estimate. We see that linear inversion produces physical estimates in all cases and there is negligible difference between the fidelities of the physical and linear inversion estimates.

\begin{table}[ht]
\caption{State Reconstruction Parameters for Different Error Types} 
\centering 
\begin{tabular}{c c c c c} 
\hline\hline 
Error & Raw Fidelity & Fidelity & Var(Fidelity) & Physicality \\ [0.5ex] 

\hline 
None & 0.8490 & 0.8491 & 0.0018 & 0\\ 
X & 0.8194  & 0.8195 & 0.0021 & 0 \\
Y & 0.8152 & 0.8148 & 0.0022 & 0 \\
Z & 0.8047 & 0.8046 & 0.0019 & 0 \\[1ex]
\hline 
\end{tabular}
\label{table:Fid}
\end{table}

\subsection*{Insensitivity to state-preparation errors}

Since we are conditioning on the measurement results of the syndrome qubits, the error-detection circuit has the useful feature that rotation errors on the prepared (encoded) two-qubit state correspond only to decreasing the success probability of preserving the desired state. From a tomographic standpoint we can accurately reconstruct the conditioned state as long as the total number of shots is large relative to the error syndrome probability so that sufficient measurement statistics are available.


\begin{suppfigure}[htbp!]
\centering
\includegraphics[width=0.47\textwidth]{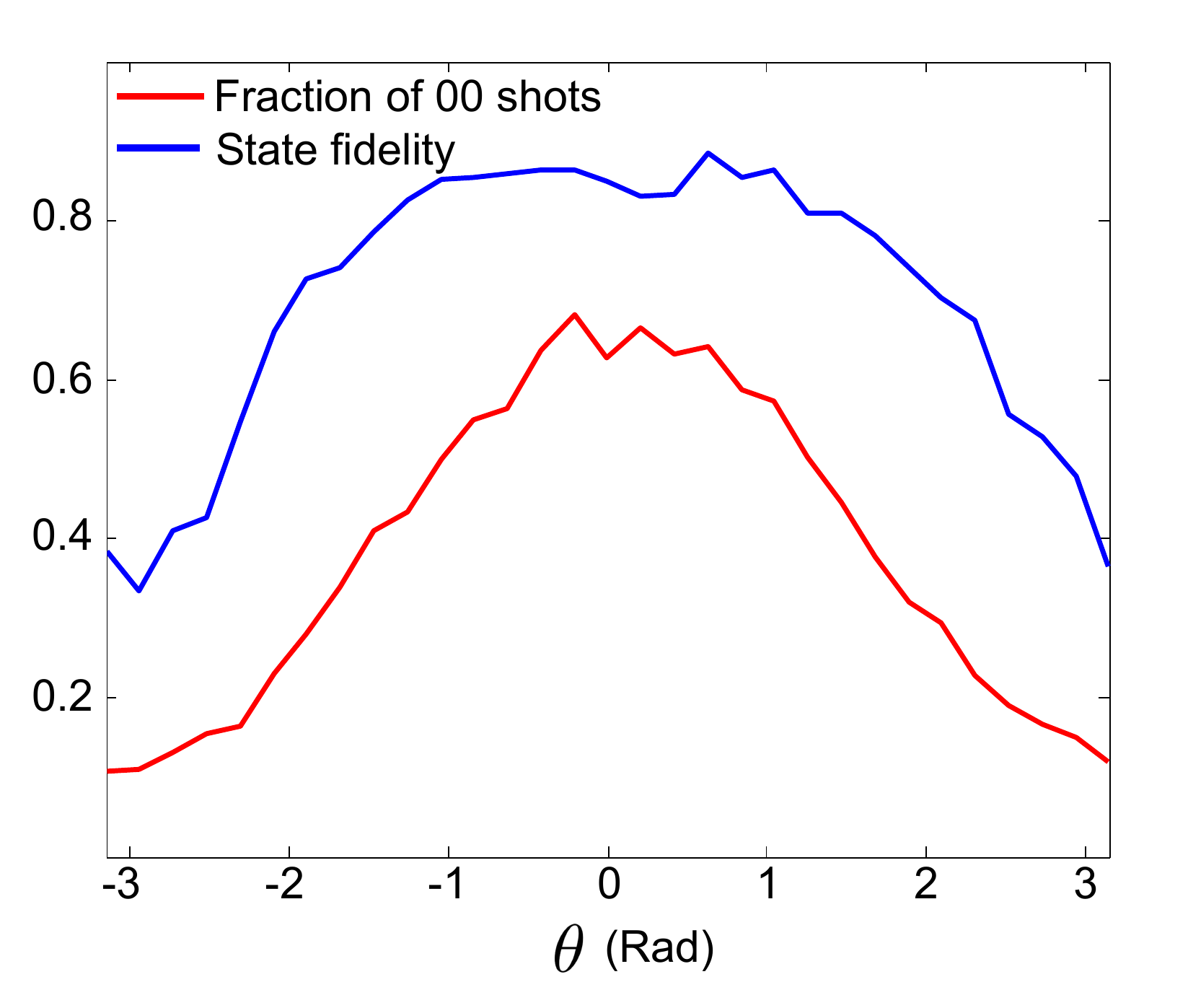}
\caption{\label{fig:Fid_and_shots_biterror} \textbf{Fidelity and Error Angle.} \textbf{a}, Fidelity and fraction of down-down shots as a function of bit error angle.}
\end{suppfigure}

To make this precise, suppose that the ideal initial state $|\psi_{00}\rangle$ is rotated via some error operator $E$ to the state
\begin{align}
|\psi_\mathrm{init}\rangle &= a|\psi_0\rangle + b|\psi_1\rangle + c|\psi_2\rangle +d|\psi_3\rangle.
\end{align}
This gives the following syndrome probabilities
\begin{align}
\text{pr}(00)&= |a|^2,\nonumber \\
\text{pr}(10)&= |b|^2,\nonumber \\
\text{pr}(01)&= |c|^2,\nonumber \\
\text{pr}(11)&= |d|^2,
\end{align}
and so the probability of successfully obtaining the correct state is $|a|^2$. Since the shots producing the error syndromes are evenly distributed throughout the different unitary rotation pulses on the code qubits, the effect on the code state is to reduce the number of shots by a factor of $|a|^2$ which can also be thought of as a re-scaling of the measurement variances by $\frac{1}{|a|^2}$. Hence, for each of the 108 different measurement observables $\mathcal{O}_j$, $\text{tr}(\rho^{00} \mathcal{O}_j)$ has a single-shot variance that scales as $\frac{V^{00}_{jj}}{|a|^2}$. This implies we expect that, to first order in $\theta$, state tomography is robust to over-under rotation errors.

We can model and verify this effect by directly applying a unitary error of varying strength on the first code qubit. The general unitary Kraus operator is $\varepsilon=\exp\left(-i\frac{\theta}{2}\hat{n}\cdot \vec{\sigma}\right)$ and the probabilities for the different syndromes are given by Eq.~(\ref{eq:error_probs}). For simplicity, we chose a purely $X$ rotation so $\varepsilon = \cos(\theta)I-i\sin(\theta)X$ and varied the size of the angle in 30 steps from $-\pi$ to $\pi$. The state fidelity $F_\text{state}(\rho^{00},|\psi^{00}\rangle\langle \psi^{00}|)$ as a function of $\theta$ is shown in Extended Data Fig.~\ref{fig:Fid_and_shots_biterror}. As expected, the first derivative appears to smoothly converge to $0$ as $\theta$ converges to $0$ and the loss in fidelity is a result of insufficient statistics for the $00$-syndrome state.

This discussion also allows us to more accurately predict the output conditional state fidelities. As demonstrated, we can effectively ignore coherent errors in the first two CNOT gates since they are used for state-preparation and errors in these operations show up as a reduction in the number of shots available for tomography. Assuming the number of shots is large enough, and ignoring single-qubit errors, we are only concerned with errors in the final three CNOT gates. From two-qubit RB the average gate fidelity of our CNOT gates is approximately 0.94. Hence, assuming depolarizing errors, we can obtain an approximate gate fidelity for the comprised circuit of $\sim 0.94^3 = 0.83$ and state fidelities with similar values, which is consistent with our obtained fidelities in Table~\ref{table:Fid}.


\smallskip\noindent
\textbf{Acknowledgments}
	We thank M.~B.~Rothwell and G.~A.~Keefe for fabricating devices. We thank J.~R.~Rozen, J.~Rohrs and K. Fung for experimental contributions. We thank S. Bravyi, and J. A. Smolin for engaging discussions. We thank I.~Siddiqi for providing the Josephson Parametric Amplifier. We acknowledge Caltech for HEMT amplifiers. We acknowledge support from IARPA under contract W911NF-10-1-0324. All statements of fact, opinion or conclusions contained herein are those of the authors and should not be construed as representing the official views or policies of the U.S. Government.

\smallskip\noindent
\textbf{Author contributions} *A.D.C., E.M., and S.J.S. contributed equally to this work.  J.M.C. and J.M.G. designed the experiments.  A.D.C. and S.J.S. characterized devices and ran the experiments. A.W.C, J.M.G., and M.S. developed the gate breakdown for the code implementation. E.M., A.D.C., and J.M.G interpreted and analyzed the experimental data. All authors contributed to the composition of the manuscript. 

\smallskip\noindent
\textbf{Author information} The authors declare no competing financial interests. Correspondence and requests for materials should be sent to A. D. C\'orcoles, adcorcol@us.ibm.com.

\clearpage

\end{document}